\documentstyle [epsbox,12pt]{article}
\def\o{\over}

\def\r{\gamma}
\def\a{\alpha}
\def\b{\beta}

\def\p{\pi}

\def\Re{{\rm Re}}
\def\bar{\overline}
\def\Im{\rm Im}

\def\t{\tilde}

\def\G{{\rm GeV}}
\begin{document}
\baselineskip=24pt
\setcounter{page}{1}
\thispagestyle{empty}
\topskip -2.5  cm
%%%%%%%%%%%%%%%%%%%%%%%%%%
\begin{flushright}
\begin{tabular}{c c}
& {\normalsize  UWThPh-1994-48}\\
& {\normalsize  AUE-08-94, US-94-06}\\
%& {\normalsize  US-94-06}\\
& {\normalsize  October 1994}
\end{tabular}
\end{flushright}
%%%%%%%%%%%%%%%%%%%%%%%%%
\vspace{0.01cm}
\centerline{\large\bf Electric Dipole Moments of Neutron and  Electron}
\centerline{\large\bf in Two-Higgs-Doublet Model with
     Maximal  $CP$ violation}
%\vskip 1 cm
\vskip 0.5 cm

\centerline{{\bf Takemi HAYASHI}$^{(a)}$, \hskip 1 cm
           {{\bf Yoshio KOIDE}$^{(b)}$
  \footnote{E-mail:koide@u-shizuoka-ken.ac.jp} }}
\centerline{{{\bf Masahisa  MATSUDA}$^{(c)}$
              \footnote{E-mail:masa@auephyas.aichi-edu.ac.jp},} \hskip 0.2 cm
           {\bf Morimitsu TANIMOTO}$^{(d)}$, \hskip 0.2 cm
           {\bf Seiichi WAKAIZUMI}$^{(e)}$}
\centerline{$^{(a)}$ \it{Kogakkan University, Ise, Mie 516, JAPAN}}
\centerline{$^{(b)}$ \it{Department of Physics, University of Shizuoka,
  52-1 Yada, Shizuoka 422, JAPAN}}
\centerline{$^{(c)}$ \it{Department of Physics and Astronomy, Aichi University
of Education}}
\centerline{\it Kariya, Aichi 448, JAPAN}
\centerline{$^{(d)}$ \it{Institut f\"ur Theoretische Physik,
               Universit\"at Wien}}
\centerline{{ \it Boltzmanngasse 5, A-1090 Wien, AUSTRIA}
  \footnote{Permanent address:Science Education Laboratory,
  Ehime   University,790 Matsuyama, JAPAN}}
\centerline{$^{(e)}$ \it{Department of Physics, School of Medical Sciences,
University of Tokushima}}
\centerline{\it Tokushima 770, JAPAN}
\vskip 0.2 cm
%\vskip 1 cm
\centerline{\bf ABSTRACT}
 We study the electric dipole moments(EDM) of the neutron and the
electron in the two-Higgs-doublet model,
in the case that $CP$ symmetry is violated maximally
in the neutral Higgs sector.
We take account of the Weinberg's operator $O_{3g}=GG\t G$
as well as the operator $O_{qg}=\bar q\sigma\t Gq$ for the neutron,
and  the Barr-Zee diagrams for the electron.
It is found that the predicted neutron EDM could be considerably reduced by
the destructive contribution of the two Higgs scalars  to get the lower
value than the experimental  bound.
As to the electron EDM, the predicted value
is smaller in one order than the experimental one.
\newpage
%%%%%%%%%%%%%%%%%%%%%%%%%%%%%%%%%%%%%%%%%%%%%%%%%%%%%%%%%%%%%%%%%%%%%%%%%%%%%%%
%%%%%%%%%%%%%%%%%%%%%%%%%%%%%%%%%%%%%%%%%%%%%%%%%%%%%%%%%%%%%%%%%%%%%%%%%%%%%%%
\topskip 1 cm
The physics of  $CP$ violation has attracted much recent attention
in the light that the $B$-factory will go on line in the near future.
The central subject of the $B$-factory is the test of the
standard  model(SM), in which the origin of $CP$
violation is reduced to the phase in the Kobayashi-Maskawa matrix[1].
However, there has been a general interest in considering other approaches
to $CP$ violation since many alternate sources exist.
The electric dipole moment(EDM) of the neutron is of central importance
to probe these new sources  of $CP$ violation, because it is very small in SM
while it can be larger in the other models.
By begining with the papers of Weinberg[2], there has been considerable renewed
interest in the EDM induced by $CP$  violating  neutral  Higgs sector.
Some studies[3,4,5] revealed numerically the importance of the
"chromo-electric" dipole moment, which arises from the three-gluon operator
$GG\t G$ found by Weinberg[2] and the light quark operator
$\bar q \sigma\t Gq$  introduced
by Gunion and Wyler[3], in the neutral Higgs sector.
We know the simplest  extension of SM Higgs sector, namely
the type II two-Higgs-doublet model(THDM)[6], which
demonstrates explicit or spontaneous $CP$ violation[7]
in its neutral sector if the soft breaking term of the discrete symmetry
is included.  Many authors have already studied this model
and proposed to search for $CP$ violating observables
directly or indirectly in the Higgs sectors[8,9,10]. \par
%%%%%%%%%%%%%%%%%%%%%%%%%%%%%%%%%%%%%%%%%%%%%%%%%%%%%
   The estimates of the maximal values of those observables are important to
search for $CP$ violating effect  in the Higgs sector.
We have already studied the Higgs potential to give
maximal $CP$ violation and its effect on  the $\rho$ parameter[11,12].
We have found that the maximal $CP$ violation is realized  under the fixed
values of  $\tan\b$ with two  constraints of parameters in the
Higgs potential[12].
The purpose  of this paper is to calculate the values of
EDM of the neutron and the electron in THDM with  maximal $CP$ violation,
taking account of the contribution of the "chromo-electric" dipole moment.
This result will give us  constraints of $CP$ violating Higgs sector by
comparing with the experimental bounds.\par
%%%%%%%%%%%%%%%%%%%%%%%%%%%%%%%%%%%%%%%%%%%%%%%%%%%%%%%%%%%%%%
First, we discuss the maximal $CP$ violation in the THDM, where
the Higgs potential with $CP$ violating terms is written as[13]:
%%%%%%%%%%%%%%%%%%%%%%%%%%%%%%%%
\begin{eqnarray}
V _{{\rm Higgs}}&=&{1\o 2}g_1(\Phi_1^\dagger\Phi_1-|v_1|^2)^2+
         {1\o 2}g_2(\Phi_2^\dagger\Phi_2-|v_2|^2)^2 \nonumber\\
  &+& g(\Phi_1^\dagger\Phi_1-|v_1|^2)(\Phi_2^\dagger\Phi_2-|v_2|^2)
      \nonumber\\
&+& g'|\Phi_1^\dagger\Phi_2-v_1^*v_2|^2+\Re[h(\Phi_1^\dagger\Phi_2-v_1^*v_2)^2]
           \nonumber\\
  &+& \xi\left [{\Phi_1\o v_1}-{\Phi_2\o v_2}\right ]^\dagger
         \left [{\Phi_1\o v_1}-{\Phi_2\o v_2}\right ] \ ,
\end{eqnarray}
%%%%%%%%%%%%%%%%%%%%%%%%%%%%%%
\noindent where
$\Phi_1$ and $\Phi_2$
couple with the down-quark and the up-quark sectors respectively
and   the vacuum expectation values are defined as
$v_1\equiv <\Phi_1^0>_{vac}$  and  $v_2\equiv <\Phi_2^0>_{vac}$.
We do not concern ourselves here with a specific model of $CP$
violation, but instead consider a general parametrization using the
notation developed by Weinberg[13].
We take the coupling constant $h$ in eq.(1) to be real and set
\begin{equation}
  v_1^* v_2=|v_1 v_2|\exp(i\phi) \ ,
\end{equation}
\noindent as a phase convension.
%%%%%%%%%%%%%%%%%%%%%%%%%%%%%%%%%%%%%%%%%%%%%%%%%%%%%%%%%%%
We define the neutral components of the two Higgs doublets
using three real fields $\phi_1,\phi_2,\phi_3$
 and the Goldstone boson $\chi^0$ as follows:
 \begin{eqnarray}
\Phi_1^0&=&{1\o\sqrt{2}}\{\phi_1+\sqrt{2}v_1+i(\cos\b\chi^0-\sin\b\phi_3)\}\ ,
        \nonumber\\
\Phi_2^0&=&{1\o\sqrt{2}}\{\phi_2+\sqrt{2}v_2+i(\sin\b\chi^0+\cos\b\phi_3)\}\ ,
\end{eqnarray}
\noindent
 where $\tan\b\equiv v_2/v_1$.
%%%%%%%%%%%%%%%%%%%%%%%%%%%%%%%%%%%%%%%%%%%%%%%%%%%%%%%%%%%
The real fields $\phi_1$ and $\phi_2$ are scalar particles while
$\phi_3$ is pseudo-scalar in the limit of $CP$ conservation.
$CP$ violation occurs via the scalar-pseudoscalar interference terms in
the neutral Higgs mass matrix.
In this basis, the mass matrix elements are given as
%%%%%%%%%%%%%%%%%%%%%%%%%%%%%%%%%%%%%%%%%%%%%%%%%%
\begin{eqnarray}
M_{11}^2&=&2g_1|v_1|^2+g'|v_2|^2+{\xi+\Re(hv_1^{*2}v_2^2)\o|v_1|^2}\ ,
\nonumber\\
M_{22}^2&=&2g_2|v_2|^2+g'|v_1|^2+{\xi+\Re(hv_1^{*2}v_2^2)\o|v_2|^2}\ ,
  \nonumber\\
M_{33}^2&=&(|v_1|^2+|v_2|^2) \left [g'+
              {\xi-\Re(hv_1^{*2}v_2^2)\o|v_1v_2|^2}\right ]
   \ ,  \nonumber\\
M_{12}^2&=&|v_1v_2|(2g+g')+{\Re(hv_1^{*2}v_2^2)-\xi\o|v_1v_2|}
     \ ,   \\
M_{13}^2&=&-{\sqrt{|v_1|^2+|v_2|^2}\o|v_1^2v_2|}\Im{\it (hv_1^{*2}v_2^2)}\ ,
\nonumber\\
M_{23}^2&=&-{\sqrt{|v_1|^2+|v_2|^2}\o|v_1v_2^2|}\Im{\it (hv_1^{*2}v_2^2)}\ ,
\nonumber
\end{eqnarray}
\noindent where the mass matrix is the symmetric one.
%%%%%%%%%%%%%%%%%%%%%%%%%%%%%%%%%%%%%%%%%%%%%%%%%%%%%%%%%%
Maximal CP violation is  defined on a new basis by Georgi[14],
where  the Goldstone boson decouples from the
 $\Phi_2$ doublet, since the gauge couplings of the Higgs bosons
 are diagonal in this basis.
The neutral Higgs scalars
$\t H^0, \t H^1, \t H^2$ on this new basis are obtained by the following
rotation;
\begin{equation}
  \left( \matrix{\t H^0 \cr \t H^1\cr \t H^2} \right )
=\left(\matrix{\cos\b& \sin\b &0\cr -\sin\b& \cos\b &0\cr 0&0&1 }  \right)
\left
   ( \matrix{\phi_1\cr \phi_2\cr \phi_3} \right ) \ .
\end{equation}
By denoting the orthogonal matrix ${\bf O}$ that relates this basis
with the mass eigenstates
$H_1$, $H_2$ and $H_3$ as
\begin{equation}
  \left( \matrix{\t H^0 \cr \t H^1\cr \t H^2} \right )
={\bf O} \left( \matrix{H_1\cr H_2\cr H_3} \right ) \ ,
\end{equation}
\noindent
  maximal $CP$ violation is defined when
\begin{equation}
  {\bf O}_{11}^2={\bf O}_{12}^2={\bf O}_{13}^2={1\o 3} \ ,
\end{equation}
\noindent
which was presented by  M\'endez and Pomarol[10].
Here, the matrix $\bf O$ is related with the orthogonal matrix ${\bf U}$, which
   is defined in ref.[12] to
diagonalize the mass matrix of eq.(4),
as follows:
\begin{equation}
  {\bf O}
=\left(\matrix{\cos\b& \sin\b &0\cr -\sin\b& \cos\b &0\cr 0&0&1 }  \right)
{\bf U}  \ .
\end{equation}

In ref.[12], the constraints of the Higgs
potential parameters to get  maximal $CP$ violation have been studied.
%%%%%%%%%%%%%%%%%%%%%%%%%%%%%%%%%
%\noindent We then have
%
%\begin{eqnarray}
%{\bf O}_{11}&=&\cos\b{\bf U}_{11}+\sin\b {\bf U}_{21}\ ,
%        \nonumber\\
%{\bf O}_{12}&=&\cos\b{\bf U}_{12}+\sin\b {\bf U}_{22}\ ,
%             \nonumber\\
%{\bf O}_{13}&=&\cos\b{\bf U}_{13}+\sin\b {\bf U}_{23}\ .
%\end{eqnarray}
%%%%%%%%%%%%%%%%%%%%%%%%%%%%%%%%%%%%%%%%%%%%%%%%%%%%%%%%%%
%%%%%%%%%%%%%%%%%%%%%%%%%%%%%%%%%%%%%%%%%%%%%%%%%%%%%%%%%%%%%
In this paper,
two solutions yielding maximal $CP$ violation, which satisfy the
condition of eq.(7), for $\tan\b$[12] have been obtained;
\begin{equation}
{\rm Sol. I}:\ \tan\b={1\o\sqrt{2}}(\sqrt{3}-1)=0.51\cdots \ ,\quad
{\rm Sol. II}:\
\tan\b={1\o\sqrt{2}}(\sqrt{3}+1)=1.93\cdots \ ,
\end{equation}
\noindent with constraints
\begin{equation}
g_1+g_2+2g-2\bar\xi=0\ ,\qquad g_1=g_2\ ,\qquad \phi={\pi\o 4}\ ,
\end{equation}
\noindent  where
${\bar \xi}\equiv \xi/|v_1v_2|^2$ and
$CP$ violating phase $\phi$ takes its maximal value as
is expected.
The condition of $g_1+g_2+2g-2\bar\xi=0$ gives an important
constraint for the neutral Higgs scalars and the charged Higgs one,
since  ${\bar \xi}$ determines the charged Higgs mass as follows:
\begin{equation}
 m^2_{H^\pm}=\bar\xi v^2 \ ,
\end{equation}
\noindent where $v^2 \equiv v_1^2+v_2^2$.
In addition to these constraints, there are the positivity conditions such
as[15]
\begin{eqnarray}
&g_1&\geq 0, \quad
g_2\geq 0, \quad
g> -\sqrt{g_1g_2}, \quad
g+g'-|h|\geq -\sqrt{g_1g_2}, \nonumber\\
&\xi&\geq 0, \quad
g'-|h|+{\bar \xi}\geq 0, \quad
{\bar \xi}-g\geq -\sqrt{g_1g_2}.
\end{eqnarray}
%%%%%%%%%%%%%%%%%%%%%%%%%%%%%%%%%%%%%%%%%%%%%%%%%%%%%
%%%%%%%%%%%%%%%%%%%%%%%%%%%%%%%%%%%%%%%%%%%%%%%
 The masses of the three neutral Higgs scalars are given as
\begin{eqnarray}
m_{H1}^2&=&2g_1\cos^4\b+2g_2\sin^4\b+4(\bar\xi-g)\sin^2\b\cos^2\b
  =2g_1  \ ,\nonumber \\
m_{H2}^2&=& g'+ \bar\xi+h \ , \qquad
m_{H3}^2= g'+ \bar\xi-h \ ,
\end{eqnarray} \noindent in units of $v^2$, where
 the  conditions in eq.(10) are used in the second equality for $m_{H1}^2$.
We notice that the four Higgs masses
are given by four parameters $g_1$, $g'$, $h$ and $\bar\xi$.
Since the parameter $h$ is predicted to be very small in
analyses using the renormalization group equation[16],
 the values of $m_{H2}$ and $m_{H3}$ are expected to be
close to each other.
In our numerical analyses, we take $m_{H2}<m_{H3}$
 by fixing  $h<0$ as our convention.  On the other hand, $m_{H1}$  is not
constrained.  \par
%%%%%%%%%%%%%%%%%%%%%%%%%%%%%%%%
%%%%%%%%%%%%%%%%%%%%%%%%%%%%%%%%%%%%%%%%%%%%%%%%%%%%
 Now, we can estimate the $CP$ violating parameters
 ${\Im} Z_1$ and ${\Im} Z_2$ in THDM.
These are  the imaginary parts of the scalar meson fields normalization
constants, $Z_i$, which are the
column vectors in the neutral Higgs scalar vector
space, defined in terms of the tree level approximation
to the two-point function as follows[13]:
%%%%%%%%%%%%%%%%%%%%%%%%%
\begin{equation}
{1 \o v_i^2}\langle \phi_i^0\phi_i^0\rangle_q
       =\sum_{n=1}^3 {\sqrt{2}G_F\o q^2-m^2_{Hn}}Z_i^{(n)} \quad(i=1,2) \ ,
\end{equation}
\noindent  where $v_i\equiv \langle \phi_i^0\rangle_{vac}$.
The $CP$ violation factors ${\rm Im} Z_i^{(n)}$ are deduced to
\begin{equation}
{\rm Im} Z_1^{(k)}=-{\tan\b\o \cos\b}u_1^{(k)} u_3^{(k)} \ , \qquad {\rm Im}
Z_2^{(k)}={1\o \tan\b\sin\b}u_2^{(k)} u_3^{(k)}\ ,
\end{equation}
\noindent
where $u_i^{(k)}$ denotes the $i-$th component of the $k-$th normalized
eigenvector of the Higgs mass matrix.
 %%%%%%%%%%%%%%%%%%%%%%%%%%%%%%%%%%%%%%%%%%%%%%%%%%%%%%%%
%%%%%%%%%%%%%%%%%%%%%%%%%%%%%%%%%%%%%%%%%%%%%%%%%%%%%%%%%%%%%%%%%%
 The values of  $u_i^{(k)}$ are given  by solving Higgs mass matrix
${\bf M^2}$ of eq.(4) in the case of the maximal $CP$ violation.
In the maximal $CP$ violation,
 we get[12]
\begin{eqnarray}
 u^{(1)}&=&(\cos\b,\ -\sin\b,\ 0 ) \ , \nonumber \\
 u^{(2)}&=&({1\o \sqrt{2}}\sin\b,\ {1\o \sqrt{2}}\cos\b,
                          \ -{1\o \sqrt{2}}) \ , \nonumber \\
 u^{(3)}&=&({1\o \sqrt{2}}\sin\b,\ {1\o \sqrt{2}}\cos\b,
                          \ {1\o \sqrt{2}}) \ .
\end{eqnarray}
\noindent Using these eigenvectors,
 we can calculate  $CP$ violating parameters
 ${\rm Im}Z_i^{(k)}$.
For the first Higgs scalar, these are zero because the third
component of the eigenvector is zero as seen in eq.(16),
i.e., there is no scalar-pseudoscalar interference term.
We have non-vanishing values for second and third Higgs scalars
(k=2,3) as follows:
\begin{equation}
  {\rm Im} Z_1^{(2)}=-{\rm Im} Z_1^{(3)}={1\o 4}(\sqrt{3}\mp 1)^2
 \ , \qquad
  {\rm Im} Z_2^{(2)}=-{\rm Im} Z_2^{(3)}=-{1\o 4}(\sqrt{3}\pm 1)^2
\ ,
\end{equation}
\noindent the upper and lower signs correspond to the Sol.I and Sol.II
 of $\tan\b$ in eq.(9), respectively.
We notice that these values are somewhat smaller than
   the Weinberg's bound[13] taking the same value of $\tan\b$,
\begin{equation}
  \left |{{\rm Im} Z_1^{(2,3)}\o {\rm Im} Z_1^{(WB)}}\right | \simeq \left\{
 \matrix {0.8
   9\cr 0.46} \right.
 \ , \qquad
 \left |{{\rm Im} Z_2^{(2,3)}\o {\rm Im} Z_2^{(WB)}}\right |\simeq
\left\{ \matrix {0.46\cr 0.89} \right.   \ ,
\end{equation}
\noindent where
$(WB)$ denotes the Weinberg's bounds, and the upper values and
lower ones correspond to the Sol.I and Sol.II of $\tan\b$, respectively.
Thus, the Weinberg's bound does not correspond to  maximal $CP$ violation,
because  both ${\rm Im} Z_1^{(k)}$ and ${\rm Im} Z_2^{(k)}$ cannot
approach to those bounds at the same time.
\par
%%%%%%%%%%%%%%%%%%%%%%%%%%%%%%%%%%%%%
%%%%%%%%%%%%%%%%%%%%%%%%%%%%%%%%%%%%%
%%%%%%%%%%%%%%%%%%%%%%%%%%%%%%%%%%%%%
  Let us discuss the EDM of the neutron
in the case of maximal $CP$ violation.
%%%%%%%%%%%%%%%%%%%%%%%%%%%%%%%%%%%%%%%%%%%%%%%%%%%%%%%%%%
The low energy $CP$ violating interaction is described by an effective
Lagrangian $L_{CP}$, which is generally decomposed into the local
composite operators $O_i$ of the quark and gluon fields,
\begin{equation}
L_{CP}=\sum_i C_i(M,\mu)O_i(\mu) \ .
\end{equation}
Some authors pointed out[2,3] that the three gluon operator with the
dimension six and the quark-gluon operator with the dimension five dominate
 EDM of the neutron in THDM. So, we study the effect of these two
operators on the neutron EDM.
 Various techniques have been developed to estimate the strong-interaction
hadronic  effects[17,18,19,20].
The simplest one is the NDA approach[17], but it provides at best the
order-of-magnitude estimates.
The systematic technique has been given by Chemtob[18] for the case of
the  operator with
 the higher-dimension involving the gluon fields.
 We employ his technique to get the hadronic matrix elements of our operators.
\par
  Let us define the following  operators:
  \begin{equation}
  O_{qg}(x)=-{g_s \o 2}\bar q\sigma_{\mu\nu}\tilde G^{\mu\nu} q \ ,\qquad
  O_{3g}(x)=-{g_s^3\o 3}f^{abc}\tilde G^{a,\mu\nu}G^{b,\a}_\mu G^c_{\nu\a} \ ,
\end{equation}
\noindent where $q$ denotes $u,d$ or $s$ quark.  The QCD corrected
coefficients are given by the two-loop calculations[2,3] as follows:
\begin{eqnarray}
 C_{ug}&=&-{\sqrt{2}G_F m_u(\mu)\o 128\p^4}g_s^2(\mu)[f(z_t)+g(z_t)]\Im Z_2
      \left ({g_s(\mu)\o g_s(M)}\right )^{-{74\o 23}} \ , \nonumber \\
 C_{dg}&=&-{\sqrt{2}G_F m_{d}(\mu)\o 128\p^4}g_s^2(\mu)
[f(z_t)\tan^2\b \Im Z_2 -{\it g(z_t)}\cot^2\b \Im Z_1]
      \left ({g_s(\mu)\o g_s(M)}\right )^{-{74\o 23}} \ ,  \nonumber\\
 C_{3g}&=&{\sqrt{2}G_F \o 256\p^4}h(z_t)\Im Z_2
      \left ({g_s(\mu)\o g_s(M)}\right )^{-{108\o 23}} \ ,
\end{eqnarray}
\noindent where $z_t=(m_t/m_H)^2$
and we omitt the upper-indices $(k)$ defined in eq.(15).
The functions $f(z_t)$, $g(z_t)$
and $h(z_t)$ are the two-loop integral functions,
which are defined in refs.[4,5,21].  The  coefficient $C_{sg}$ is
obtained from $C_{dg}$ by repalcing $m_d$ with $m_s$.\par
The hadronic matrix elements of the two operators
 are approximated by the intermediate states with the single nucleon pole
 and the  nucleon plus one pion. Then, the nucleon matrix elements are  defined
   as[18]
\begin{eqnarray}
  \langle N(P)|O_i(0)|N(P)\rangle &=& A_i\bar U(P)i\r_5 U(P) \ , \nonumber\\
  \langle N(P')|O_i|N(P)\p(k)\rangle &=& B_i\bar U(P')\tau^a U(P) \ ,
\end{eqnarray}
\noindent
where $U(P)$ is the normalized nucleon Dirac spinors
               with the four momuntum $P$.
By using $A_i$ and $B_i(i=ug,dg,sg,3g)$, the neutron EDM, $d_n^\r$,
is written as
\begin{equation}
  d_n^\r={e\mu_n\o 2 m_n^2}\sum_i C_i A_i +
  F(g_{\p NN},m_n,m_\p )\sum_i C_i B_i    \ ,
\end{equation}
\noindent where $\mu_n$ is the neutron anomalous magnetic moment.
 The function $F(g_{\p NN},m_n,m_\p )$ was derived by calculating the pion
and nucleon loop corrections using the chiral Lagrangian
for the coupled $N\p\r$ and is given in Appendix A of ref.[18].
Here, the dimensional regularization with the  standard $\bar {MS}$ scheme
is  used for defining the finite parts of the divergent integrals.
The coefficients $A_i$ and $B_i$ were given by the use of the large $N_c$
current algebra and the $\eta_0$ meson dominance[18].
Then,  we have
\begin{equation}
 A_i=f_i g_{\eta_0 NN} \ ,\hskip 1.5 cm B_i=-{4(m_u+m_d)a_1f_i\o F_\p F_0} \ ,
\end{equation}
\noindent with $a_1=-(m_{\Sigma^0}-m_{\Sigma})/(2m_s-m_u-m_d)\simeq -0.28$
and $F_\p=\sqrt{2/3}F_0=0.186\G$,
 where  $f_i$ is defined by
\begin{equation}
 \langle \eta_0(q)| O_i(0)|0 \rangle \equiv f_i q^2 \ .
\end{equation}
\noindent
 The values of  $f_i$ were derived by using QCD sum rules as follows[18]:
\begin{equation}
 f_{qg}=-0.346\G^2 \ , \hskip 2 cm f_{3g}=-0.842 \G^3\ ,
\end{equation}
\noindent
 where $f_{qg}$ denotes the flavor singlet coupling.
\par
%%%%%%%%%%%%%%%%%%%%%%%%%%%%%%%%%%%%%%%%%%%%%%%%%%%%%
Now, we can calculate the neutron EDM. Our input parameters
are[20]
\begin{eqnarray}
& &\Lambda_{QCD}=0.26\G \ ,  \qquad (m_u,m_d,m_s)=(5.6,9.9, 200) {\rm MeV}\ ,
   \qquad    \mu=m_n \ ,    \nonumber \\
& & M=m_t=150 \G \ , \qquad  g_{\p NN}=13.5 \ , \qquad g_{\eta_0 NN}=0.892 \ .
\end{eqnarray}
We show in fig.1 our predictions of the neutron EDM versus
 the ratio of the masses of the two Higgs scalars in the case of
$m_{H2}=200,\ 400,\ 600\G$ by using Sol.I of the maximal $CP$ violation.
 The predicted neutron EDM in Sol.II is almost same as the one in
Sol.I.
 As seen in fig.1, the mass difference  $\Delta m_H=m_{H3}-m_{H2}$ should be
small in order to get the lower values of EDM than the experimental upper
bound, $11\times 10^{-26}e\cdot$cm[22].
We get $\Delta m_H \simeq 18\G$, which corresponds to  $|h|\simeq 0.13$ ,
in the case of $m_{H2}=200\G$,
$\Delta m_H \simeq 52\G$($|h|\simeq 0.73$)  in the case of $m_{H2}=400\G$, and
$\Delta m_H \simeq 140\G$($|h|\simeq 3.12$)  in the case of $m_{H2}=600\G$.
In order to show the effect of the relevant operators,
 we show the contributions of $O_{ug}$, $O_{dg}+O_{sg}$ and $O_{3g}$ to the
neutron  EDM in fig.2 in the case of $m_{H2}=200\G$ by using Sol.I.
  As seen in fig.2,  the effects of the
$O_{ug}$ and $O_{3g}$ cancel out each other, and the $O_{dg}+O_{sg}$
 contribution dominates the neutron EDM, which has already been pointed out
 in ref.[20].  Since the contributions
of  $O_{ug}$ and $O_{3g}$ depend only on $\Im {\it Z_2}$ as seen in eq.(21),
there are large difference between Sol.I and Sol.II
 in these contributions.
However, these contributions almost cancel out
each other in the predicted total neutron EDM.
 On the other hand, the contribution of $O_{dg}+O_{sg}$ in Sol.I
 is exactly same as the one in Sol.II.  Thus, both Sol.I and Sol.II
 give the almost same values for the predicted neutron EDM.
 %%%%%%%%%%%%%%%%%%%%%%%%%%%%%%%%%%%%%%%%%%%%%%%%%%
\begin{figure}
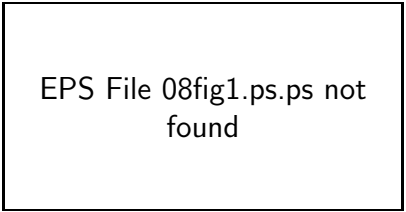

\begin{center}
\epsfile{file=08fig1.ps,scale=0.7}
\end{center}
\caption{The predicted neutron EDM versus the ratio of
 $m_{H2}/m_{H3}$ in the cases of
 $m_{H2}=200\G$(solid curve), $400\G$(dashed curve) and
 $600\G$(dashed-dotted curve) by using Sol.I.
The horizontal dashed line denotes the experimental upper bound,
$11\times 10^{-26}e\cdot{\rm cm}$.}
\end{figure}
%%%%%%%%%%%%%%%%%%%%%%%%%%%%%%%%%%%%%%%%%%%%%%%%%%%%%

 Now let us discuss the EDM of the electron.
Barr and Zee[5] presented the two-loop Feynman diagrams which can
lead to a large EDM of the charged lepton.
Those diagrams involve a heavy particle, say the top quark or
$W$ boson in the loop
that couples to an external photon line as follows:
\begin{eqnarray}
\left[{d_e\o e}\right ]_{t-loop}\ &=&-{\a\o {12\p^3}}\sqrt{2}G_Fm_e
   \left[ f(z_t)\tan^2\b{\Im}Z_2+g(z_t)\cot^2\b{\Im}Z_1\right ]
  \ ,   \\
 \left[{d_e\o e}\right ]_{W-loop}&=&-{\a\o {32\p^3}}\sqrt{2}G_Fm_e
   \left[ 3f(z_W)+5g(z_W)\right ](\sin^2\b \tan^2\b{\Im}Z_2 +\cos^2\b{\Im}Z_1)
   \ , \nonumber
\end{eqnarray}
\noindent where $z_t=(m_t/m_H)^2$ and $z_W=(m_W/m_H)^2$.
The loop-functions $f(z)$ and $g(z)$ are the same ones as those in eq.(21).
 Chang, Keung and Yuan[8] have given the complete set of two-loop diagrams
in the multi-Higgs-doublet model.
They calculate the effective  $CP$ violating $HZ\r$ vertex
 in addition to  the $H\r\r$ one
induced  by the unphysical charged Higgs and the
  $W$ contribution.
In this paper, we use their results[8] instead of eq.(28), however
 the structures of $\tan\b$ and ${\Im} Z_i$ in eq.(28) being unchanged.
%%%%%%%%%%%%%%%%%%%%%%%%%%%%%%%%%%%%%%%%%%%%%%%%%%%%%%%%%%%%%%%%%%%%%%
\begin{figure}
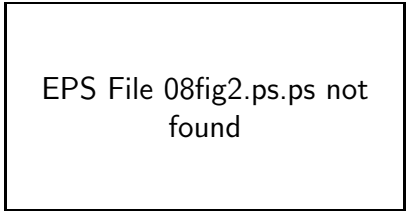

\begin{center}
\epsfile{file=08fig2.ps,scale=0.7}
\end{center}
\caption{The components of the predicted neutron
EDM versus the ratio
of  $m_{H2}/m_{H3}$ in the cases of
 $m_{H2}=200\G$ by using Sol.I.  The thick-solid curve denotes the absolute
 values of the total contribution. The thin-solid curve,
 the dashed curve and the dashed-dotted curve denote
 the contribution from $O_{ug}$, $O_{dg}+O_{sg}$ and $O_{3g}$
 operators, respectively.}
\end{figure}
%%%%%%%%%%%%%%%%%%%%%%%%%%%%%%%%%%%%%%%%%%%%%%%%%%%%%%

  In the case of maximal $CP$ violation, the loop contributions
 of $W$  and the unphysical charged Higgs boson
  vanish because $\sin^2\b \tan^2\b{\Im} Z_2 +\cos^2\b {\Im} Z_1$
 is zero in both Sol.I and Sol.II.  So, the elctron EDM
 is given by only the top-quark loop contribution, which is
 exactly same in both Sol.I and Sol.II.
 We show in fig.3 the predicted electron EDM versus the ratio of the two
Higgs scalars in the case of $m_{H2}=200,\ 400, \ 600\G$. Thus, the
magnitude of  the electron EDM
 does not go over the experimental  bound
$(-0.3\pm 0.8)\times 10^{-26}e\cdot$cm[22]
even if the mass difference of the
two Higgs scalars is considerably large.
%%%%%%%%%%%%%%%%%%%%%%%%%%%%%%%%%%%%%%%%%%%%%%%%%%
\begin{figure}
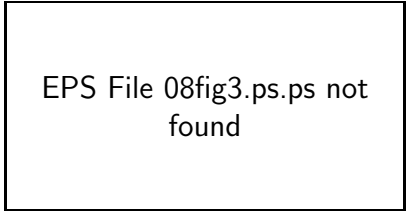

\begin{center}
\epsfile{file=08fig3.ps,scale=0.7}
\end{center}
\caption{The predicted electron EDM versus the ratio of
 $m_{H2}/m_{H3}$ in the cases of
 $m_{H2}=200\G$(solid curve), $400\G$(dashed curve) and
 $600\G$(dashed-dotted curve).
The experimental upper bound $(-0.3\pm 0.8)\times 10^{-26}e
\cdot {\rm cm}$
is  outside  the figure.}
\end{figure}
%%%%%%%%%%%%%%%%%%%%%%%%%%%%%%%%%%%%%%%%%%%%%%%%%%
 Summary is given as follows.
 We have studied the EDM of the neutron and the electron
 in the two Higgs doublet model, in the case of $CP$ symmetry being
 violated maximally
in the neutral Higgs sector.
The maximal $CP$ violation is realized  under the fixed values
of  $\tan\b$ with  two  constraints of parameters in the Higgs potential.
  We have taken account of the Weinberg's operator $O_{3g}=GG\t G$
and the operator $O_{qg}=\bar q\sigma\t Gq$ for the neutron,
 and  Barr-Zee diagrams for the electron.
 It is found that the predicted neutron EDM could be considerably reduced
due to the destructive contribution of the two Higgs scalars leading to
the lower value than the experimental
 upper  bound, while the predicted electron EDM  is smaller in one order
   than the experimental bound.
 Since our predicted value of the neutron EDM lies around the present
 experimental bound,  we expect its experimental improvement to reveal
 the new physics beyond SM.
\par
%%%%%%%%%%%%%%%%%%%%%%%%%%%%%%%%%%%%%%%%%%%%%%%%%%%%
\vskip 0.5 cm
\centerline{\bf Acknowledgments}\par
One of us(M.T) thanks the particle group at Institut f\"ur Theoretische Physik
   in Universit\"at Wien, especially,
Prof.H. Pietschmann, for kind hospitality.
This research is supported  by the Grant-in-Aid for Scientific
Research, Ministry of Education, Science and Culture, Japan(No.06220101 and
No.06640386).
%%%%%%%%%%%%%%%%%%%%%%%%%%%%%%%%%%%%%%%%%%%%%%%%%%%%
%%%%%%%%%%%%%%%%%%%%%%%%%%%%%%%%%%%%%%%%%%%%%%%%%%%%%%%
\newpage
\topskip -0.7 cm
\centerline{{\large \bf References}}
\vskip 0.3 cm
\noindent
[1] M.Kobayashi and T.Maskawa, Prog. Theor. Phys. {\bf 49}(1973) 652.\par
\noindent
[2] S. Weinberg, Phys. Rev. Lett. {\bf 63}(1989)2333.\par
\noindent
[3] J.F. Gunion and D. Wyler, Phys. Letts. {\bf 248B}(1990)170.\par
\noindent
[4] A. De R\' ujula, M.B. Gavela, O. P\` ene and F.J. Vegas,
    Phys. Lett. {\bf 245B}(1990)640; \par
    N-P. Chang and D-X. Li, Phys. Rev. {\bf D42}(1990)871;\par
  D.Chang, T.W.Kephart, W-Y.Keung and T.C.Yuan,
                    Phys.Rev.Lett. {\bf 68}(1992)439;\par
 M. J. Booth and G. Jungman, Phys. Rev. {\bf D47}(1993)R4828.\par
\noindent
[5] S.M. Barr and A. Zee, Phys. Rev. Lett. {\bf 65}(1990)21;\par
   S.M. Barr, Phys. Rev. Lett. {\bf 68}(1992)1822;
            Phys. Rev. {\bf D47}(1993)2025.\par
\noindent
[6] For a text of Higgs physics see J.F. Gunion, H.E. Haber, G.L.Kane
and S. Dawson,\par
  {\it "Higgs Hunter's Guide"},  Addison-Wesley, Reading, MA(1989). \par
\noindent
[7] G.C. Branco and M.N. Rebelo, Phys. Lett. {\bf 160B}(1985)117.
\par\noindent
[8] J.F. Gunion and R. Vega, Phys. Lett. {\bf 251B}(1990)157;\par
   D. Chang, W-Y. Keung and T.C. Yuan,  Phys. Rev. {\bf   D43}(1991)14;\par
  R.G. Leigh et al., Nucl. Phys. {\bf B352}(1991)45.\par
\noindent
[9] B. Grz\c adkowski and J.F. Gunion, Phys. Lett.
                            {\bf 287B}(1992)237;\par
  B. Grz\c adkowski and J.F. Gunion, Phys. Lett.
                            {\bf 294B}(1992)361;\par
  B. Grz\c adkowski and W-Y. Keung, Phys. Lett.
                            {\bf 319B}(1993)526;\par
  C. Schmidt and M. Peskin, Phys. Rev. Lett. {\bf 69}(1992)410;\par
R. Cruz, B. Grz\c adkowski and J.F. Gunion, Phys. Lett.
                            {\bf 289B}(1992)440;\par
D. Atwood, G. Eilam and A. Soni, Phys. Rev. Lett.
       {\bf 70}(1993)1364;\par
 B. Grz\c adkowski , preprint at
        Warsaw University, IFT07/94(1994),hep-ph@9404330.\par
 \noindent
[10] A. M\'endez and A. Pomarol, Phys. Lett.
                            {\bf 272B}(1991)313.\par
 \noindent
[11] C.D. Froggatt, R.G. Moorhouse  and I.G. Knowles,
   Phys. Rev. {\bf D45}(1992)2471;\par
   Nucl. Phys. {\bf B386}(1992)63;\par
   A. Pomarol and R. Vega, Nucl.Phys. {\bf B413}(1994)3. \par
 \noindent
[12] G.C. Joshi, M. Matsuda and M. Tanimoto, UWThPh-1994-25, AUE-04-94, \par
   hep-ph@9407255(1994), to be published in Phy. Lett. {\bf B}. \par
 \noindent
[13] S. Weinberg, Phys. Rev. {\bf D42}(1990)860.\par
\noindent
[14]  H. Georgi, Hadr. J. Phys. {\bf 1}(1978)155.\par
\noindent
[15] B. Kastening, Private communications and see the preprint(
hep-ph@9307224).\par
\noindent
[16] M.A. Luty, Phys. Rev. {\bf D41}(1990)2893;\par
     C.D. Froggatt, I.G. Knowles, R.G. Moorhouse,
     Phys. Lett. {\bf 249B}(1990)273; \par
  M. Chemtob, Z. Phys. {\bf C60}(1993)443.\par
 \noindent
[17] A. Manohar and H. Georgi, Nucl. Phys. {\bf B234}(1984)189.\par
\noindent
[18] M. Chemtob, Phys. Rev. {\bf D45}(1992)1649.\par
\noindent
[19]X-G. He, B.H.J. Mckellar and S. Pakvasa, Mod. Phys. {\bf A4}(1989)5011;\par
  I.I. Bigi and N.G. Uraltsev, Nucl. Phys. {\bf B353}(1991)321.\par
\noindent
[20] T.Hayashi, Y.Koide, M.Matsuda and M.Tanimoto,
          Prog.Theor.Phys.{\bf 91}(1994)915.\par
 \noindent
[21] D.A. Dicus, Phys. Rev. {\bf D41}(1990)999;\par
  D. Chang, W-Y. Keung and T.C. Yuan, Phys. Lett. {\bf 251B}(1990)608.\par
\noindent
[22] Particle Data Group, Phys. Rev. {\bf D50}(1994)1218.\par
\end{document}